\documentstyle[11pt,aas2pp4]{article}

\begin{document}

\def\sec{$^{\prime\prime}$}
\def\min{$^{\prime}$}

\hyphenation{a-ni-so-tro-pic flux---ca-li-bra-ted}

\title{A Comparison of Radio Axis with Host Galaxy Plane Axis in Seyfert
Galaxies}

\author{Henrique R. Schmitt \altaffilmark{1,2,4},
Anne L. Kinney\altaffilmark{1}, Thaisa Storchi-Bergmann\altaffilmark{2}
and Robert Antonucci\altaffilmark{3}}

\altaffiltext{1}{Space Telescope Science Institute, 3700 San Martin Drive,
Baltimore, MD21218} 
\altaffiltext{2}{Departamento de Astronomia, IF-UFRGS, CP 15051, CEP91501-970,
Porto Alegre, RS, Brazil}
\altaffiltext{3}{University of California Santa Barbara, Physics Department,
Santa Barbara, CA93106}
\altaffiltext{4}{CNPq Fellow\\ 
email:schmitt@stsci.edu, kinney@stsci.edu, thaisa@.if.ufrgs.br, 
antonucci@physics.ucsb.edu}

\begin {abstract}

We use the radio axis as an indicator of the orientation of the obscuring torus
in Seyfert galaxies, and
analyze the difference between the position angles of extended radio
structures and host galaxy major axis of Seyfert 1 and Seyfert 2 galaxies.
We find that Seyfert 1's are
less likely to have extended radio structures along the host galaxy major axis,
while Seyfert 2's have these structures distributed in most directions.
We also find a zone of avoidance in the distribution of position angles;
both Seyfert 1's and Seyfert 2's seem to avoid close alignment between the radio
axis and the host galaxy plane axis. These results are
analyzed from the point of view of a model in which Seyfert 1's have their
obscuring torus axis aligned preferentially along the host galaxy disk axis,
and Seyfert 2's have their torus axis laying at an intermediate angle between
the galaxy disk and its axis.

\end{abstract}


\newpage

\section{Introduction}

Since the discovery by Antonucci \& Miller (1985) of polarized broad emission
lines in the nuclear spectrum of NGC1068, the interest in the Unified Model
for Seyfert galaxies has grown considerably (see Antonucci 1993 and Urry \&
Padovani 1995 for a review of the model). This model assumes that
Seyfert 1's and Seyfert 2's both contain an active nucleus surrounded by a dusty
torus, and that the angle through which the central engine is observed determines
the classification of the object. Several pieces of
evidence corroborate this model,
like the observation of polarized broad emission lines in several Seyfert 2
galaxies (Miller \& Goodrich 1990), deficit of ionizing photons in Seyfert 2's
(Wilson, Ward \& Haniff 1988; Kinney et al. 1991) and the collimation of the
nuclear radiation, observed as extended linear radio sources (Ulvestad \& Wilson
1989) and conically shaped Narrow Line Regions (Pogge 1989).

In a recent paper Schmitt \& Kinney (1996) studied the NLR shape of Seyfert
galaxies, as predicted by the Unified Model, using archival high resolution HST
[OIII] images. Their results show that the sizes of Seyfert 1's NLR's are much
smaller than the NLR's of Seyfert 2's would be
if they were observed pole-on. The sample has not been selected by an isotropic
property, but the similarity in radio and $\lambda$5007\AA\ luminosities of
Seyfert 1's and Seyfert 2's suggests the comparison may be rather fair anyway.
This result apparently contradicts the Unified Model, from
which we would expect both kinds of objects to have similar intrinsic NLR sizes. 
In order to solve this problem, Schmitt \& Kinney (1996)
propose that the Seyfert 1's torus axis may be
aligned preferentially along the host galaxy plane axis, while the Seyfert 2's
torus axis may lie at an intermediate angle between the galaxy plane axis and
the galaxy plane. In this picture, because
the amount of extended gas intercepted by the nuclear radiation is
smaller perpendicular to the plane than at directions closer to it, the NLR
appears more extended in Seyfert 2's than in Seyfert 1's. 

The reason for the
Seyfert 1 orientation distribution may be that even those objects which are seen
from the polar nuclear torus direction, but nearly edge-on to the host galaxy are
classified as Seyfert 2 because of obscuration by
dust in the host galaxy plane. There is
ample evidence that the broad line region
in edge-on galaxies is reddened (de Zotti \&
Gaskell 1985) or entirely extinguished (Keel 1980; Lawrence \& Elvis 1982).

In order to check the orientation of the torus axis in Seyfert 1's and
Seyfert 2's, in this paper we compare the position angle (PA)
of the extended nuclear radio structures of Seyfert 1's and Seyfert 2's
with their host galaxy's
major axis PA. We show that there is a lack of
Seyfert 1's with radio structures aligned
along the host galaxy major axis (which would correspond to the case where the 
obscuring torus is nearly perpendicular to
the galaxy plane), while Seyfert 2's have 
radio structures distributed along almost all directions, consistent with the
model proposed by Schmitt \& Kinney (1996).

\section{The Sample and Measurements}

We selected from the literature a sample of 46 Seyfert galaxies (15 Seyfert 1's
and 31 Seyfert 2's) having high resolution radio maps and showing
linear or slightly resolved radio structures, as defined by Ulvestad \& Wilson
(1984a). In Table 1 we give the galaxy names, together with their
activity type (Seyfert 1 or Seyfert 2), Morphological Type, the mean numerical
index (T) of stage along the Hubble sequence (as defined in
de Vaucouleurs et al. 1991), distance, 6cm 
flux, logarithm of the 6 cm power, PA of the radio structure (PA$_{RAD}$), 
PA of major axis (PA$_{MA}$), the difference between 
PA$_{RAD}$ and PA$_{MA}$ ($\Delta$PA) and the inclination of the host galaxy. 
The distances were calculated using the
galaxy's radial velocities relative to the local group (de Vaucoulers et al.
1991) and H$_0=$75 km s$^{-1}$ Mpc$^{-1}$. The inclinations were
calculated from the axial ratios, assuming that the face-on galaxy is basically
circular in shape.

The PA's of the extended radio emission were obtained from the references
in Table 1, using their published values 
or measuring it on their radio maps when the PA was not
given explicitly. The host galaxies major axis PA's were mostly obtained from
de Vaucoulers et al. (1991), with the remainder from
references cited in Table 1.
The PA's were checked by looking at the galaxies on the
Digitized Sky Survey Plates. 
For the galaxies without values for the major axis PA
available in the literature and for the cases where the published value was
wrong (NGC5929, MRK573, MCG-8-11-11),
the PA was measured by fitting ellipses over the outer isophotes of
the digitized galaxy image.

\section{Results}

Figure 1 shows the histogram of the differences between the PA's of the radio
axis and of the host galaxy major axis.
We can see that there is a lack of
small values for Seyfert 1's, indicating that they are less likely to have 
extended radio structures along the host galaxy major axis. In other
words, the cases in which the radio axis lies in the plane of the host galaxy
are rare in Seyfert 1's, as expected.
Meanwhile, the Seyfert 2's have $\Delta$PA values evenly distributed from 
0$^{\circ}$
to 70$^{\circ}$. However, it is interesting to note that
both distributions have a lack of objects with
radio structures well aligned with the host galaxy minor axis.
Applying a KS test to the data, we find that the hypothesis
that the two groups of galaxies are drawn from the same parent population is
rejected at the 99.0\% level, or 99.1\% when we exclude double nuclei
galaxies (NGC1144, MRK110, MRK266 and MRK463E).

We must ask if the Seyfert 1's and 2's in the sample have similar intrinsic  
properties, or if the above results might be traceable
to selection effects.
First we compare the logarithm of the 6cm radio power of the two groups, to
see if they are similar in the two groups of galaxies, since
the more luminous objects might be expected to have larger radio structures, which
are easier to detect and measure. In Figure 2 we show the histogram of the
logarithm of the 6cm radio power, where we can see that both
groups have very similar distributions of radio
powers, except for the high luminosity tail of Seyfert 2's.
This high luminosity tail had already been observed by Meurs \& Wilson (1984)
and Wilson \& Ulvestad (1989). However, these objects are undistinguished in 
the position angle histograms, so they do not create the claimed effects.
Note that the former reference includes disk
emission which may be significant in low luminosity objects. 

Wilson \& Tsvetanov (1994) have recently proposed that the obscuring
torus axis may be aligned with the galaxy plane axis in late type galaxies,
while the obscuring torus
could have any orientation in early type systems. In order to check
if our sample is biased towards Seyfert 1's in late type galaxies
and Seyfert 2's in early type galaxies, 
we show in Figure 3 a histogram of their morphological types, where the
parameter T$=-$4 corresponds to Ellipticals, the earliest morphological type,
and T$=$5 corresponds to Sc's, the latest morphological type.
The group distributions are almost equal,
with the only difference being the existence of some Seyfert 2's with T$\leq-$2
(as early as S0)
without any corresponding Seyfert 1's in this region of the diagram. 
These objects cannot be distinguished in the position angle histograms,
assuring us that
the groups are not biased by different morphological types.

We also check for a trend for the obscuring torus axis to
align along the minor axis in late type galaxies. In Figure 4 we show a
plot of the  morphological type versus $\Delta$PA.
There is no systematic trend for late type objects to
have large $\Delta$PA values, which means that their radio structures are not
preferentially aligned with the host galaxy minor axis.
This result suggests that the trend
found by Wilson \& Tsvetanov (1994) was most likely due to the small number of
objects in their sample.

\section{Discussion}

The results presented in Figure 1 can be interpreted from the point of view 
of the scheme proposed by Schmitt \& Kinney (1996), where the Seyfert 1's have
their torus axis aligned preferentially close to the galaxy plane axis, while
the Seyfert 2's have their torus axis aligned at an intermediate angle between
the galaxy plane and its axis. Here we describe a simple model,
developed in order to study the orientation of the projected radio
structure (assumed to be aligned with the torus axis) due
to its inclination relative to the line of sight.
In our model we assume a uniform distribution of orientation angles between the 
obscuring torus axis and the host galaxy plane axis. Then we predict what the
observed values for $\Delta$PA would be, given that uniform distribution.

Figure 5 shows a cartoon which represents our model.
In this Figure, the galaxy disk lies on the X-Y plane and the
elongated radio structure,
represented by the unit vector $\hat{\hbox{\bf k}}_j$, 
makes an angle $\beta$ with the 
Z-axis (galaxy plane axis). The elongated radio structure,
is equally likely to be in any position along
the circle ``c'', which is described by the azimuthal angle $\theta$, 
measured from the
X to the Y-axis. The galaxy is inclined relative to the line of sight
(Z$^{\prime}$-axis) by an angle ``i'', 
such that the inclined galaxy major axis is
coincident with the X-axis.

In this model the angle $\beta$, 
between the radio axis and the galaxy plane
axis assumes values in the range
1$^{\circ}\leq\beta\leq$90$^{\circ}$ and is varied in steps of 1$^{\circ}$. For 
each value of $\beta$ we generate one
thousand vectors equally spaced along one half the circle ``c''. 
Due to the symmetry of the problem, the angle $\theta$ is varied only in the range 
--90$^{\circ}\leq\theta\leq$90$^{\circ}$. 
Also, due to the fact that the circle ``c'' describes different perimeters for
different values of $\beta$, 
the results were weighted for each value of $\beta$,
by the area of a ring of 1$^{\circ}$ described by the circle ``c'' on 
the surface of a sphere,
relative to the whole sphere area. This is necessary in order to avoid an 
oversampling in the number of vectors for 
smaller $\beta$ values, relative to
larger ones.
The galaxy inclination ``i'' is then varied in the range 
15$^{\circ}\leq$i$\leq$60$^{\circ}$ in steps of 1$^{\circ}$. We chose
this range because for i$<$15$^{\circ}$ it 
is difficult to measure the galaxy inclination
angle and it is considered to be face on, while for i$>$60$^{\circ}$, there
is only a small number of galaxies with such inclination in our sample. 
The small number of objects with i$>60^{\circ}$ suggests an inclination
dependent selection bias.
We have also tested the model for smaller and larger values of ``i'', but
the final result does not change.

The projected components of each individual radio jet
vector (for every $\theta$, $\beta$ and i value), in the
directions X, Y$^{\prime}$ and Z$^{\prime}$,
can be calculated by the following relations:

$$P_X=\cos \theta \sin \beta$$

$$P_{Y^{\prime}}=\cos i \sin \theta \sin \beta + \sin i \cos \beta$$
 
$$P_{Z^{\prime}}=\cos i \cos \beta - \sin i \sin \theta \sin \beta$$

\noindent
The observed difference between
the elongated radio structure
PA and the major axis PA ($\Delta$PA) is given by:

$$\Delta PA=\arctan (P_{Y^{\prime}}/P_X)$$

\noindent
We divide $\Delta$PA in 10$^{\circ}$ bins and count the number of
vectors with projected
$\Delta$PA inside each bin, for a given $\beta$ value. This number, normalized
to the total number of vectors, can be
considered as the probability of an elongated 
radio structure, which makes an angle $\beta$ with
the host galaxy axis, being seen at a projected angle $\Delta$PA from the
galaxy major axis, considering inclination effects.

The model also includes some constraints. First, a galaxy is only considered 
to be a Seyfert 1 if the angle between 
Z$^{\prime}$ and $\hat{\hbox{\bf k}}_j$ is less than
30$^{\circ}$, corresponding to observing into the opening of the torus;
otherwise it is a Seyfert 2. This number was obtained from 
Osterbrock \& Shaw (1988) and corresponds to half the torus opening angle. 
Second, if the projected length (r) of the unit vector $\hat{\hbox{\bf k}}_j$
in the X-Y$^{\prime}$ plane
is less than a given value, it is assumed to be unresolved and is not
considered when we count the number of
vectors that fall inside different $\Delta$PA bins. 
This corresponds to the case where the elongated
radio structure is observed pole on, and is
therefore not observed as elongated. Since we do not know the intrinsic length
of the elongated radio structure, 
we note that they usually have sizes of
the order of 1\arcsec--2\arcsec\ and that the VLA resolution at 6cm in A
configuration is $\approx$0.4\arcsec. We thus
study two cases: r$>$0.2 and r$>$0.4.
Notice also, that Seyfert 2's are not effected
by the projected size restriction, because the vectors with angle
between Z$^{\prime}$ and $\hat{\hbox{\bf k}}_j$
larger than 30$^{\circ}$ will have r$>$0.5.

Figure 6 shows the model results. The individual lines are cumulative
histograms, representing the sum over
all $\beta$ angles smaller than or equal to the value indicated 
at the right and below each line.
The vertical axis of these plots were
normalized relative to the $\Delta$PA bin
with the largest number of vectors on the $\beta=$90$^{\circ}$ bin. 
The histogram can be
considered as the cumulative probability of an elongated radio structure, 
that makes an angle
smaller than or equal to $\beta$, being seen at a projected angle $\Delta$PA
from the galaxy major axis.
In Figure 6a we show the results for Seyfert 1's with projected 
length (r) larger than 0.4. Figure 6b shows
the results for Seyfert 1's with projected length larger than 0.2 and figure 6c
shows the results for Seyfert 2's. The $\Delta$PA's
distribution for
Seyfert 1's are very similar, independent of the ``r'' value. For small
$\beta$ values there is a larger probability of observing an elongated radio
structure closer to the host galaxy minor axis (large $\Delta$PA). 
For larger $\beta$ values, the probability of observing
the elongated radio structure 
closer to the major axis increases. When we consider the distribution
for all $\beta$ angles together (the 90$^{\circ}$ line),
the probability of finding a vector at any $\Delta$PA bin 
is approximately the same. 
The distribution of $\Delta$PA's for Seyfert 2's is similar to
that of Seyfert 1's. However, the probability of finding small $\Delta$PA's,
which corresponds to an elongated radio structure close to the host galaxy major
axis, only increases when we go to larger $\beta$ values relative to those
of Seyfert 1's.

Comparing the model results from Figure 6 with the observed $\Delta$PA's
in Figure 1, we see that the lack of Seyfert 1 galaxies with small
$\Delta$PA values can only be understood if there are very few Seyfert 1
galaxies in which the angle between the elongated radio structure
and the galaxy plane axis ($\beta$ in the model) is large 
($\beta>30^{\circ}-40^{\circ}$). Similarly, when we
compare the results for the Seyfert 2 models with the observed values,
the distribution of $\Delta$PA values can be understood if the Seyfert 2
galaxies can take on any value for the angle
between the elongated radio structure and the galaxy plane axis.

These results support the scheme proposed by Schmitt \& Kinney (1996).
However, that scheme cannot explain the lack of both 
Seyfert 1's and Seyfert 2's with 
large $\Delta$PA values, which would correspond to elongated
radio structures aligned with the host galaxy minor
axis. This lack of large $\Delta$PA values suggests a physical
zone of avoidance, already observed by Ulvestad \& Wilson (1984b) on
a smaller sample, where
for some reason the radio axis is not ever closely aligned intrinsically
with the galaxy
plane axis. If we consider a flat distribution of $\Delta$PA's, we
calculate that for our sample of 46 galaxies, we
would expect 10 galaxies with $\Delta$PA$>70^{\circ}$. Considering
a sample of only 31 galaxies, which corresponds to the number of Seyfert 2's
in our sample, we would expect 7 galaxies with $\Delta$PA$>70^{\circ}$.
From Poisson statistics we calculate that the a posteriori
probability of observing
only one galaxy with $\Delta$PA$>70^{\circ}$, when the expected number is
10, is 10$^{-4.78}$, or 10$^{-3.47}$ for the case when 7 galaxies are expected.

There may be some
effect, which we are not taking into account, that makes large
$\Delta$PA unobservable. One 
possible solution for the lack of Seyfert 1's with elongated radio structures 
aligned with the host galaxy minor axis could be that when the obscuring torus
is closely aligned with the host galaxy axis, an elongated radio structure 
is not produced. Due to the low HI
density in the central region of spiral galaxies (Rots 1975, Begeman 1988),
the radio ``jet'' would not interact with an enough large quantity
of matter and consequently would not radiate enough to be
detected. Alternatively, these nuclear disks in the same plane as the host
galaxy may have fewer dissipation$/$fueling mechanisms. Yet another possibility
is that the nuclear axes reflect a past triggering merger, and that the approaches
of companions are statistically anisotropic.
To incorporate such effects, we show in Figure 7a and b (r$>$0.4 and r$>$0.2
respectively)
the models for Seyfert 1's with $\beta\leq$10$^{\circ}$ excluded. In this case
the probability of observing an elongated radio structure
with $\Delta$PA$>$70$^{\circ}$ is smaller
than for values of $\Delta$PA$<70^{\circ}$. 
This makes the observed and modeled $\Delta$PA
distributions look much more alike, although we would still expect 
to detect some objects with $\Delta$PA$>$70$^{\circ}$.
While this solution can perhaps solve the problem for Seyfert 1's,
it does not suffice for Seyfert 2's, where the statistics are better. A simple 
visual inspection of Figure 6c shows that the exclusion of all vectors with 
$\beta\leq10^{\circ}$ does not change the distribution of $\Delta$PA's 
significantly. Even if we exclude all vectors with $\beta\leq$30$^{\circ}$, the
probability of observing a Seyfert 2 with $\Delta$PA$>$70$^{\circ}$ would be
smaller, but it would not explain the deficit that we observe.

The lack of Seyfert 1 galaxies with elongated
radio structures
aligned with the galaxy major axis is consistent
with results obtained by
Keel (1980) and Maiolino \& Rieke (1995). These authors
showed that there is a deficiency of edge-on Seyfert 1 galaxies and
that Seyferts 1.8 and 1.9 are more likely to be in edge-on galaxies.
These results suggest that in the case of edge on objects we can be
observing the nuclear region directly, through the polar region of the nuclear
torus, but not seeing the broad line region due to shadowing by
gas and dust in the galaxy disk.

On the other hand, our results differ from those from Colbert et al.
(1996) and Baum et al. (1993). These authors found that the large scale
($\approx$1 kpc) radio structure of Seyferts 
are {\it preferentially} aligned with the host
galaxy minor axis. Baum et al. (1993) have also compared the small scale
radio structures with
the large scale radio structures and found that their PA's are different.
Colbert et al. (1996) suggest that the small scale radio jets are
possibly diverted by dense molecular clouds, like the scenario
proposed for NGC1068 by Gallimore, Baum \& O'Dea (1996), and then generate
the large scale radio structures that we see.

\section{Summary and Conclusions}

We have shown in this Paper that there is a lack of Seyfert 1 galaxies with
radio structures aligned with the host galaxy major axis, while for
Seyfert 2's the radio structures are oriented along almost
any direction in the galaxy. Both groups also show a deficiency of objects with
elongated radio structures closely aligned with the host galaxy plane axis. 

We developed a model to calculate the distribution of $\Delta$PA, the difference
between the orientation of the
radio axis and host galaxy major axis, based on the assumption that the
angle between the radio axis and the host galaxy plane axis ($\beta$) is uniformly
distributed between 0$^{\circ}$ and 90$^{\circ}$,
taking into account the effect of the galaxy 
inclination and resolution of the elongated radio structure. 
From the comparison of the observed 
$\Delta$PA distribution of Seyfert 1's with the distribution predicted by the 
model, we can explain the small number of objects with small 
$\Delta$PA's only if their torus axis lies closer than $\approx30^{\circ}$ to
the host galaxy axis. The lack of Seyfert 1's with $\Delta$PA$>70^{\circ}$ can
be partially explained if we assume that the elongated
radio structures are not often produced closer than
$\approx10^{\circ}$ to the host galaxy plane axis, and several possible
explanations for this are suggested.
For the  Seyfert 2's, the observed distribution can be explained
if the torus axis assumes any angle relative to the host galaxy plane, with the
exception of the region with $\Delta$PA$>70^{\circ}$.

These results, together with those from Keel (1980) and Maiolino \& 
Rieke (1995), showing that Seyfert 1's are less likely to be found in edge-on
galaxies, as well as the results from Lawrence \& Elvis (1982) and deZotti 
\& Gaskell (1985), showing that the broad line region in edge-on galaxies
is reddened, are in good agreement with the model proposed by Schmitt \& Kinney
(1996). The paucity of objects showing close alignment between radio axis
and galaxy plane axis remains unexplained.

\acknowledgements

This work was supported by NASA under grant 
AR-05810.01-94A and by the Brazilian institution
CNPq. This research has made use of the NASA$/$IPAC Extragalactic Database
(NED) which is operated by the Jet Propulsion Lab, Caltech, under contract 
with NASA. We also would like to thank E. Colbert for useful discussions.

\clearpage

\begin{figure}
\caption{Histogram of the difference between the radio and the host galaxy
major axis position angles of Seyfert 1's (dashed line) and Seyfert 2's
(solid line).}
\end{figure}

\begin{figure}
\caption{The distribution of the logarithm of the 6cm radio power. 
Lines as in Figure 1.}
\end{figure}

\begin{figure}
\caption{The distribution of the galaxies morphological types (T). Lines as
in Figure 1. T$=-$4 corresponds to morphological type E, T$=-$2 corresponds
to S0, T$=$0 corresponds to S0$/$a, T$=$2 corresponds to Sab and T$=$4
corresponds to Sbc.}
\end{figure}

\begin{figure}
\caption{Comparison between the radio axis and major axis position angle 
differences and the host galaxy Morphological Type. Open circles are Seyfert 1's
and filled squares are Seyfert 2's.}
\end{figure}

\begin{figure}
\caption{This cartoon shows our radio ``jet'' model. The host galaxy disk lies
in the X-Y plane, the radio ``jet'' is represented by the vector 
$\hat{\hbox{\bf k}}_j$, which
makes an angle $\beta$ with the Z-axis (galaxy plane axis)
and is likely to be in any direction
along the circle ``c'', which is described by the angle $\theta$. The angle
``i'' describes the galaxy inclination relative to the line of sight.}
\end{figure}

\begin{figure}
\caption{Results of the models. The lines are the cumulative sum of all the
vectors with $\beta$ less and equal the value indicated bellow the line. 
a) Sy1's with r$>$0.4; b) Sy1's with r$>$0.4;
c) Sy2's}
\end{figure}

\begin{figure}
\caption{Results of the Sy1 models for the case when we exclude the cases with
$\beta$ less and equal to 10$^{\circ}$. a) r$>$0.4; b) r$>$0.2.}
\end{figure}

\end{document}


\begin{deluxetable}{lrlrrrcrrrrr}
\tablewidth{0pc}
\tablecaption{Sample Properties}
\tablehead{\colhead{Name}&\colhead{Type}&\colhead{Morphological}&\colhead{T}&
\colhead{Dist.}&\colhead{S$_{6cm}$}&\colhead{Log(P$_{6cm}$)}&\colhead{PA$_{RAD}$}&
\colhead{PA$_{MA}$}&\colhead{$\Delta$PA}&\colhead{i}&\colhead{Ref.}\cr
&&Type&&(Mpc)&(mJy)&(W Hz$^{-1}$)&$\deg$&$\deg$&$\deg$&$\deg$&}
\startdata
NGC591&2&(R')SB0$/$a&0.0$\pm$0.8&62.6&7.9&21.57&$-$28&5&33&40&2,11,14\cr
NGC1068&2&(R)SA(rs)b&3.0$\pm$0.3&15.3&1090.0&22.49&28&70&42&32&1,1,14\cr
NGC1144&2&Ring B&$-$&115.3&29.3&22.67&62&130&68&$-$&9,10,14\cr
NGC1386&2&SB(s)o&$-$0.6$\pm$0.5&9.9&13.0&20.18&45&25&20&68&1,13,14\cr
NGC2210&2&SAB0&$-$3.0$\pm$0.8&28.7&175.0&22.24&1&$-$18&19&40&1,1,15\cr
NGC2273&2&SB(r)a&0.5$\pm$0.5&25.7&19.0&21.18&90&50&40&39&1,1,14\cr
NGC2992&1.9&Sa pec&1.0$\pm$0.3&28.3&77.0&21.87&$-$20&6&26&72&1,1,1\cr
NGC3227&1.5&SAB(s)pec&1.0$\pm$0.3&14.4&34.0&20.93&173&155&18&48&1,10,14\cr
NGC3362&2&SABc&5.0$\pm$0.8&109.3&2.5&21.55&57&90&33&38&9,10,14\cr
NGC3516&1.5&(R)SB(s)0&$-$2.0$\pm$0.4&36.7&4.3&20.84&20&55&35&40&1,12,10\cr
NGC4051&1.5&SAB(rs)bc&4.0$\pm$0.3&10.2&6.0&19.87&81&135&54&41&1,1,14\cr
NGC4151&1.5&(R')SAB(rs)ab&2.0$\pm$0.3&13.6&125.0&21.44&77&26&51&44&1,10,17\cr
NGC4117&2&S0&$-$2.3$\pm$0.3&12.2&0.6&19.03&0&18&18&60&2,2,14\cr
NGC4388&2&SA(s)b:sp&3.0$\pm$0.4&32.8&76.0&22.00&23&92&69&77&1,1,14\cr
NGC5135&2&SB(l)ab&2.0$\pm$0.3&52.8&58.8&22.29&25&$-$5&30&46&2,2,18\cr
NGC5252&1.9&S0&$-$2.0$\pm$0.8&91.4&13.5&22.13&$-$10&5&15&50&8,10,15\cr
NGC5273&1.9&SA(s)o&$-$2.0$\pm$0.3&14.9&0.9&19.38&0&10&10&27&1,1,14\cr
NGC5548&1.5&(R')SA(s)0$/$a&0.0$\pm$0.4&69.2&15.0&21.94&160&125&35&22&6,1,15\cr
NGC5643&2&SAB(rs)c&5.0$\pm$0.3&14.2&20.0&20.68&87&92&5&30&7,7,18\cr
NGC5728&2&(R l)SAB(r)a&1.0$\pm$0.4&36.5&4.6&20.87&$-$53&30&83&55&2,8,14\cr
NGC5929&2&Sab:pec&2.0$\pm$0.7&35.8&24.7&21.58&62&45&17&26&1,1,18\cr
NGC7172&2&Sa pec sp&1.4$\pm$0.5&35.0&1.8&20.42&90&100&10&56&5,5,14\cr
NGC7450&1.5&(R)SB(r)a&1.0$\pm$0.8&43.7&1.7&20.59&100&35&65&31&1,1,1\cr
NGC7672&2&Sb&3.0$\pm$0.9&57.0&1.0&20.59&95&50&45&41&2,2,18\cr
IC4329A&1&SA0 pec&$-$0.7$\pm$0.8&62.2&24.0&22.05&95&45&50&73&5,5,14\cr
ESO428-G14&2&SA:(l)0&$-$1.6$\pm$0.6&19.0&28.2&21.09&129&135&6&55&2,11,14\cr
MCG8-11-11&1.5&S&$-$&82.5&71.0&22.76&127&180&53&44&4,1,18\cr
MRK3&2&S0&$-$2.0$\pm$1.1&55.0&361.0&23.12&86&30&56&27&3,1,15\cr
MRK6&1&SAB0&$-$0.5$\pm$0.6&79.3&100.0&22.88&177&130&47&51&3,1,14\cr
MRK34&2&S&$-$&196.8&6.1&22.45&$-$22&27&49&27&3,1,1\cr
MRK78&2&S0$/$E&$-$3.5$\pm$1.5&150.5&12.0&22.51&90&85&5&60&3,1,1\cr
MRK79&1.2&SBb&3.0$\pm$0.8&89.1&4.2&21.60&2&65&63&46&3,1,16\cr
MRK110&1&Pair&$-$&137.5&2.2&21.70&88&50&38&$-$&3,3,18\cr
MRK176&2&SA(s)0$/$a pec&1.0$\pm$0.9&109.4&10.4&22.17&90&55&35&63&3,3,18\cr
MRK266&2&Compact pec&$-$&112.6&22.4&22.71&32&7&25&$-$&9,10,18\cr
MRK268&2&S&$-$&156.7&17.3&22.71&70&95&25&48&3,1,1\cr
MRK270&2&S0&$-$2.0$\pm$1.7&37.9&5.7&21.00&48&114&66&25&1,1,15\cr
MRK279&1&S0&$-$2.0$\pm$0.4&124.3&7.7&22.15&90&33&57&56&3,3,14\cr
MRK348&2&SA(s)0$/$a&0.0$\pm$0.4&62.3&480.0&23.35&170&111&59&27&3,1,15\cr
MRK463E&2&Pair&$-$&202.8&100.0&23.69&0&70&70&$-$&3,1,10\cr
MRK509&1&Compact&$-$&135.1&4.5&22.00&110&75&35&25&3,3,18\cr
MRK533&2&SA(r)bc pec&4.0$\pm$0.5&118.8&66.5&23.05&117&147&30&25&9,10,18\cr
MRK573&2&(R)SAB(rs)0&$-$1.0$\pm$0.5&68.8&5.5&21.49&125&180&55&44&3,1,18\cr
MRK618&1&SB(s)b pec&3.0$\pm$0.9&143.1&4.2&22.01&146&80&66&39&3,3,18\cr
MRK766&1.5&(R')SB(s)a&1.0$\pm$0.6&50.9&15.1&21.67&16&66&50&37&3,3,15\cr
MRK1066&2&(R)SB(s)0&$-$1.0$\pm$0.8&49.4&35.5&22.02&135&90&45&54&2,11,14\cr
\tablenotetext{}{We consider Seyferts 1, 1.2 and 1.5 as Seyfert 1's and
Seyferts 1.8, 1.9 and 2.0 as Seyfert 2's}
\tablenotetext{}{The Morphological Types were obtained
from de Vaucoulers et al. (1991), except MRK78 (Wilson \& Tsvetanov 1994)}
\tablenotetext{}{Column 12 gives the references for S$_6$, PA$_{RAD}$ and
PA$_{MA}$, respectively. 1) Ulvestad \& Wilson 1984b, 
2) Ulvestad \& Wilson 1989, 3)Ulvestad \& Wilson 1984a, 4) Ulvestad, Wilson \&
Sramek 1981, 5) Unger et al. 1987, 6) Wilson \& Willis 1980, 7) Morris et al.
1985, 8) Wilson \& Tsvetanov 1994, 9) Edelson 1987, 10) Kukula et al. 1995,
11) Bower et al. 1995, 12) Miyaji, Wilson \& P\'erez-Fournon 1992,
13) Weaver, Wilson \& Baldwin 1991, 14) de Vaucouleurs et al. 1991,
15) Mulchaey, Wilson \& Tsvetanov 1996, 16) Oke \& Lauer 1979, 17) Simkin 1975,
18) This paper}
\enddata
\end{deluxetable}